\documentclass[twoside,twocolumn,9pt]{article}
\usepackage{extsizes}
\usepackage[super,sort&compress,comma]{natbib} 
\usepackage[version=3]{mhchem}
\usepackage[left=1.5cm, right=1.5cm, top=1.785cm, bottom=2.0cm]{geometry}
\usepackage{balance}
\usepackage{mathptmx}
\usepackage{sectsty}
\usepackage{graphicx} 
\usepackage{lastpage}
\usepackage[format=plain,justification=justified,singlelinecheck=false,font={stretch=1.125,small,sf},labelfont=bf,labelsep=space]{caption}
\usepackage{float}
\usepackage{fancyhdr}
\usepackage{fnpos}
\usepackage[english]{babel}
\addto{\captionsenglish}{%
  
}
\usepackage{array}
\usepackage{droidsans}
\usepackage{charter}
\usepackage[T1]{fontenc}
\usepackage[usenames,dvipsnames]{xcolor}
\usepackage{setspace}
\usepackage[compact]{titlesec}
\usepackage{hyperref}
\usepackage{color,soul}

\usepackage{booktabs}
\usepackage{multirow}
\usepackage{tikz}

\usepackage{epstopdf}

\definecolor{cream}{RGB}{222,217,201}

\usepackage{amssymb}
\usepackage{pifont}
\newcommand{\cmark}{\ding{51}}%
\newcommand{\xmark}{\ding{55}}%

\usepackage{xcolor}

\begin{document}

\pagestyle{fancy}
\thispagestyle{plain}
\fancypagestyle{plain}{
\renewcommand{\headrulewidth}{0pt}
}

\makeFNbottom
\makeatletter
\renewcommand\LARGE{\@setfontsize\LARGE{15pt}{17}}
\renewcommand\Large{\@setfontsize\Large{12pt}{14}}
\renewcommand\large{\@setfontsize\large{10pt}{12}}
\renewcommand\footnotesize{\@setfontsize\footnotesize{7pt}{10}}
\makeatother

\renewcommand{\thefootnote}{\fnsymbol{footnote}}
\renewcommand\footnoterule{\vspace*{1pt}%
\color{cream}\hrule width 3.5in height 0.4pt \color{black}\vspace*{5pt}} 
\setcounter{secnumdepth}{5}

\makeatletter 
\renewcommand\@biblabel[1]{#1}            
\renewcommand\@makefntext[1]%
{\noindent\makebox[0pt][r]{\@thefnmark\,}#1}
\makeatother 
\renewcommand{\figurename}{\small{Fig.}~}
\sectionfont{\sffamily\Large}
\subsectionfont{\normalsize}
\subsubsectionfont{\bf}
\setstretch{1.125} 
\setlength{\skip\footins}{0.8cm}
\setlength{\footnotesep}{0.25cm}
\setlength{\jot}{10pt}
\titlespacing*{\section}{0pt}{4pt}{4pt}
\titlespacing*{\subsection}{0pt}{15pt}{1pt}

\fancyfoot{}
\fancyfoot[LO,RE]{\vspace{-7.1pt}\includegraphics[height=9pt]{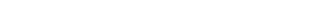}}
\fancyfoot[CO]{\vspace{-7.1pt}\hspace{13.2cm}\includegraphics{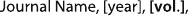}}
\fancyfoot[CE]{\vspace{-7.2pt}\hspace{-14.2cm}\includegraphics{head_foot/RF}}
\fancyfoot[RO]{\footnotesize{\sffamily{1--\pageref{LastPage} ~\textbar  \hspace{2pt}\thepage}}}
\fancyfoot[LE]{\footnotesize{\sffamily{\thepage~\textbar\hspace{3.45cm} 1--\pageref{LastPage}}}}
\fancyhead{}
\renewcommand{\headrulewidth}{0pt} 
\renewcommand{\footrulewidth}{0pt}
\setlength{\arrayrulewidth}{1pt}
\setlength{\columnsep}{6.5mm}
\setlength\bibsep{1pt}

\makeatletter 
\newlength{\figrulesep} 
\setlength{\figrulesep}{0.5\textfloatsep} 

\newcommand{\topfigrule}{\vspace*{-1pt}%
\noindent{\color{cream}\rule[-\figrulesep]{\columnwidth}{1.5pt}} }

\newcommand{\botfigrule}{\vspace*{-2pt}%
\noindent{\color{cream}\rule[\figrulesep]{\columnwidth}{1.5pt}} }

\newcommand{\dblfigrule}{\vspace*{-1pt}%
\noindent{\color{cream}\rule[-\figrulesep]{\textwidth}{1.5pt}} }

\makeatother

\twocolumn[
  \begin{@twocolumnfalse}
{\includegraphics[height=30pt]{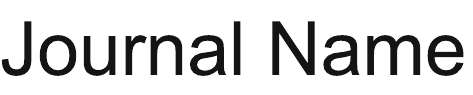}\hfill\raisebox{0pt}[0pt][0pt]{\includegraphics[height=55pt]{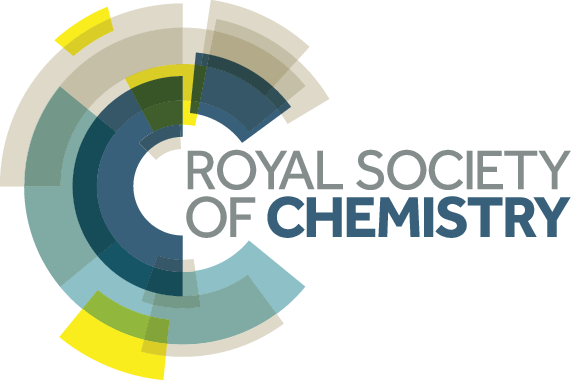}}\\[1ex]
\includegraphics[width=18.5cm]{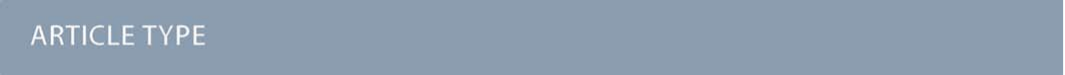}}\par
\vspace{1em}
\sffamily
\begin{tabular}{m{4.5cm} p{13.5cm} }

\includegraphics{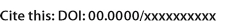} & \noindent\LARGE{\textbf{Exploring the Expertise of Large Language Models in Materials Science and Metallurgical Engineering$^\dag$}} \\
\vspace{0.3cm} & \vspace{0.3cm} \\

 & \noindent\large{Christophe BAJAN\textit{$^{a, \ddag}$} and Guillaume LAMBARD\textit{$^{a,\ddag}$}} \\

\includegraphics{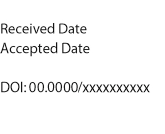} & \noindent\normalsize{The integration of artificial intelligence into various domains is rapidly increasing, with Large Language Models (LLMs) becoming more prevalent in numerous applications. This work is include in an overall project which aims to train an LLM specifically in the field of materials science. To assess the impact of this specialized training, it is essential to establish the baseline performance of existing LLMs in materials science. In this study, we evaluated 15 different LLMs using the MaScQA question answering (Q\&A) benchmark. This benchmark comprises questions from the Graduate Aptitude Test in Engineering (GATE), tailored to test models' capabilities in answering questions related to the materials science and metallurgical engineering. Our results indicate that closed-source LLMs, such as Claude-3.5-Sonnet and GPT-4o, perform the best with an overall accuracy of $\rm \sim 84$\%, when the open-source models, Llama3-70b and Phi3-14b, top at $\rm \sim 56$\% and $\rm \sim 43$\%, respectively. These findings provide a baseline for the raw capabilities of LLMs on Q\&A tasks applied to materials science, and emphasise the substantial improvement that could be brought to open-source models via prompt engineering and fine-tuning strategies. We anticipate that this work could push the adoption of LLMs as valuable assistants in materials science, demonstrating their utilities in this specialised domain and related sub-domains.} \\

\end{tabular}

 \end{@twocolumnfalse} \vspace{0.6cm}
] 

\renewcommand*\rmdefault{bch}\normalfont\upshape
\rmfamily
\section*{}
\vspace{-1cm}


\footnotetext{\textit{$^{a}$~Data-Driven Material Design Group, National Institute for Materials Science, Tsukuba, Japan. E-mail: BAJAN.Christophe@nims.go.jp, LAMBARD.Guillaume@nims.go.jp}}

\footnotetext{\dag~Supplementary Information available: [details of any supplementary information available should be included here]. See DOI: 00.0000/00000000.}

\footnotetext{\ddag~These authors contributed equally to this work}



\section{Introduction}

Large Language Models (LLMs) represent a significant advancement in artificial intelligence (AI), demonstrating exceptional proficiency in natural language processing (NLP). These models are designed to generate human-like text based on the patterns extracted from large pre-training data. LLMs have shown notable progress in a range of NLP tasks, including text generation, translation, summarization, and question answering on various benchmarks. 

However, LLMs' capabilities often degrade when addressing domain-specific requests, such as those in materials science~\cite{ref1}. This limitation arises because pre-training data typically comes from diverse web sources, encompassing a wide range of domains. While this approach effectively compresses general knowledge into the LLM's parameters, it can lead to the merging of unrelated contexts during inference, potentially resulting in incorrect assertions.

To overcome this challenge and effectively utilize LLMs for domain-specific tasks, two primary strategies can be employed:

\begin{enumerate}
    \item[(i)] Train a dedicated LLM from scratch with a smaller parameter count, specifically tailored to encapsulate the desired domain knowledge.
    \item[(ii)] Fine-tune a pre-trained LLM to a specific domain~\cite{lu2024fine}.
\end{enumerate}

In this study, we adopt the second strategy, leveraging the instruction-following capabilities and general NLP proficiency of pre-existing models. Our final objective is to fine-tune an existing LLM and integrated it into a retrieval-augmented generation (RAG) system for materials science applications. To guide this future fine-tuning process and establish a baseline for evaluation, we first assess in the present study the capabilities of available LLMs in materials science. This evaluation aims to:

\begin{itemize}
    \item Establish a comprehensive baseline performance on materials science tasks.
    \item Identify LLMs that balance high capabilities with modest parameter counts, crucial for efficient fine-tuning and deployment.
    \item Discover potential areas for improvement in the evaluation process itself.
\end{itemize}

\subsection{LLMs in Materials Science}

Recent years have witnessed significant advancements in leveraging LLMs for materials science and engineering. Domain-specific models and tools have emerged to address the challenges of applying NLP techniques to scientific research. Notable examples include:

\begin{itemize}
    \item MatBERT~\cite{walker2021impact,TREWARTHA2022100488}: A BERT-based model fine-tuned on materials science literature, enabling tasks such as information extraction and text classification.
    \item Mat2Vec~\cite{Tshitoyan2019}: Provides word embeddings tailored for materials science, facilitating semantic analysis and knowledge representation.
    \item KGQA4MAT~\cite{an2024knowledgegraphquestionanswering}: A knowledge-based system demonstrating the utility of knowledge graph question answering for structured scientific reasoning, particularly in applications like metal-organic frameworks.
    \item HoneyComb~\cite{zhang-etal-2024-honeycomb}: Highlights the adaptability of LLMs to specialized agent-based systems that can assist in materials research workflows.
\end{itemize}

Furthermore, frameworks like SciQAG~\cite{wan2024sciqagframeworkautogeneratedscience} have been developed to automatically generate question-answer (Q\&A) pairs from scientific literature, addressing the need for domain-specific Q\&A datasets. These efforts complement existing benchmarks such as ChemLLMBench~\cite{guo2023largelanguagemodelschemistry} (for chemistry), MultiMedQA~\cite{MultiMedQA} (for medicine), and SciEval~\cite{sun2023scieval} (for STEM domains).

Despite these advancements, there remains a need for tailored benchmarks that specifically evaluate LLMs' understanding of materials science concepts. The MaScQA benchmark~\cite{ref1} addresses this gap by providing a curated dataset of 650 questions covering diverse sub-fields within materials science, including thermodynamics, atomic structure, mechanical behavior, and materials characterization. It allows for evaluating fundamental comprehension, conceptual reasoning, and numerical problem-solving—capabilities essential for real-world materials science tasks.

\subsection{The MaScQA Benchmark}

While MaScQA is the most comprehensive benchmark tailored specifically to materials science and metallurgical engineering, alternative Q\&A datasets focus on related scientific domains:

\begin{itemize}
    \item SciQ~\cite{SciQ}: A general science dataset with 13,679 questions across physics, chemistry, and biology, useful for evaluating broader scientific reasoning.
    \item ChemData700k and ChemBench4k~\cite{zhang2024chemllm}: Benchmarks designed for chemistry competency, focusing on tasks related to chemical properties, reactions, and structures.
    \item MoleculeQA~\cite{lu2024moleculeqa}: A dataset for molecular-level reasoning, particularly useful for tasks involving molecular properties and design.
    \item Custom Datasets~\cite{savit2023domain}: Researchers have also created domain-specific datasets from materials science literature by combining manual and automated generation approaches.
\end{itemize}

These alternatives offer valuable insights but either lack the specificity of MaScQA or focus on narrower aspects of chemistry and molecular properties. MaScQA remains unique in its ability to test both conceptual understanding and numerical reasoning across diverse materials science sub-fields, making it the most suitable benchmark for this study.

Originally consisting of 650 questions derived from the Graduate Aptitude Test in Engineering (GATE), the MaScQA benchmark was refined by ourselves by manually removing 6 Q\&A samples due to issues such as duplication or missing information (see Table 1 in the supplementary materials for details). This minor reduction does not significantly bias the evaluation outcomes.

The MaScQA benchmark is categorized by four types of questions:

\begin{itemize}
    \item 283 Multiple Choice Questions (MCQ)
    \item 70 Matching Type Questions (MATCH)
    \item 67 Numerical Questions with Multiple Choices (MCQN)
    \item 224 Numerical Questions (NUM) 
\end{itemize}

\begin{figure*}[!t]
\centering
  \includegraphics[width=\textwidth]{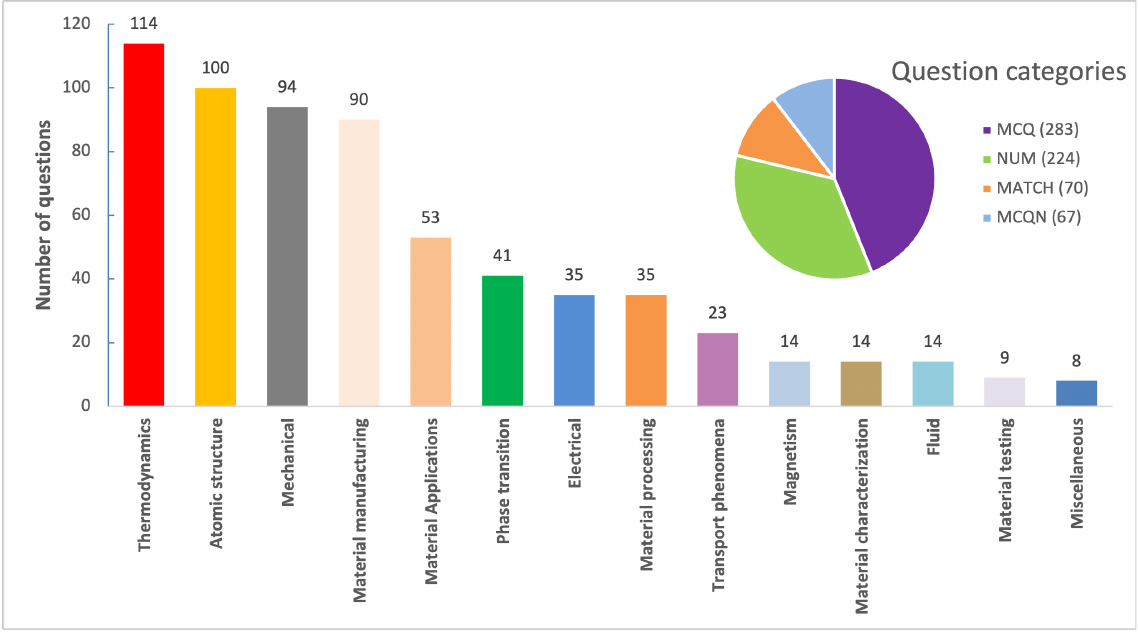}
  \caption{Distribution of the number of questions per sub-field. On the top-right hand, the number of questions per type is also reported. Figure updated from Zaki~\textit{et al.}~\cite{ref1} after removal of 6 Q\&A samples from the original MaScQA dataset.}
  \label{fgr:distribution}
\end{figure*}

These question types test various aspects of materials science knowledge, from conceptual understanding to numerical problem-solving. The questions span 14 distinct sub-fields within materials science, as shown in Figure~\ref{fgr:distribution}. 

We selected this benchmark due to its comprehensive coverage of various domains within materials science, the substantial number of questions with answers curated by hand by the MaScQA authors, and the diversity of question types that necessitate both broad knowledge and computational abilities. By establishing a baseline of LLM performance on the MaScQA benchmark, we can better understand their current limitations and potential areas for improvement in materials science applications.

\subsection{LLM Selection}

The selection of LLMs for this study encompasses a diversity of closed- and open-source models listed in Table~\ref{tbl:Information}. This diversity ensures a comprehensive evaluation across different architectures, accessibility, and fine-tunability~\cite{chen2023matchat, xie2023darwin}. The models were sourced from leading AI research organizations and companies, including Anthropic, OpenAI, Meta, Mistral AI, and Microsoft.

By evaluating models from these varied sources, we aim to capture a broad spectrum of performance characteristics, enabling a more thorough understanding of the current state of LLMs applied to materials science. This approach allows us to assess not only the raw performance of these models in answering materials science questions but also to capture the trade-off between their accessibility, affordability, and customization potential for further domain-specific fine-tuning~\cite{chiang2024llamp, song2023matsci}.

The choice of LLMs reflects models that were widely used and publicly available at the time of experimentation. Including both older and newer versions of the same models (e.g., GPT-3.5-turbo and GPT-4) enables us to track progress and evaluate incremental improvements in reasoning and performance for domain-specific tasks. While newer models, such as Llama 3.1, were released after our experiments, the results presented here provide a valuable baseline for future comparisons. Notably, improvements observed for Llama 3.1:70b on benchmarks like MATH~\cite{hendrycks2021measuringmathematicalproblemsolving} suggest that further evaluation on MaScQA could yield insightful comparisons.

\begin{table}[h]
\small
  \caption{\ List of the LLMs and their characteristics selected for this study.}
  \label{tbl:Information}
  \begin{tabular*}{0.48\textwidth}{@{\extracolsep{\fill}}llccc}
    \toprule
    \textbf{Models}     & \textbf{Developer}    & \textbf{\shortstack{Open-source}}     & \textbf{\shortstack{Fine-\\tuning}}    & \textbf{\shortstack{Number of \\ Parameters}}\\
    \midrule
    Claude-3-Haiku      & Anthropic             & \xmark                                         & \xmark                                     & --- \\
    Claude-3-Opus       & Anthropic             & \xmark                                         & \xmark                                     & --- \\ 
    Claude-3.5-Sonnet   & Anthropic             & \xmark                                         & \xmark                                     & --- \\
    GPT-3.5-turbo       & OpenAI                & \xmark                                         & \cmark                            & --- \\
    GPT-4               & OpenAI                & \xmark                                         & \cmark                                      & --- \\
    GPT-4-turbo         & OpenAI                & \xmark                                         & \xmark                                     & --- \\
    GPT-4o              & OpenAI                & \xmark                                         & \cmark                                      & --- \\
    GPT-4o-mini         & OpenAI                & \xmark                                         & \cmark                            & --- \\
    Llama2-7b           & Meta                  & \cmark                                & \cmark                            & 7B    \\
    Llama2-70b          & Meta                  & \cmark                                & \cmark                            & 70B   \\
    Llama3-8b           & Meta                  & \cmark                                & \cmark                            & 8B    \\
    Llama3-70b          & Meta                  & \cmark                                & \cmark                            & 70B   \\
    Mistral-7b          & Mistral AI            & \cmark                                & \cmark                            & 7B    \\
    Phi3-3.8b           & Microsoft             & \cmark                                & \cmark                            & 3.8B  \\
    Phi3-14b            & Microsoft             & \cmark                                & \cmark                            & 14B   \\
    \bottomrule
  \end{tabular*}
\end{table}

\section{Methodology}

\subsection{LLMs preparation}
    Our study diverges from the original work from Zaki~\textit{et al.}~\cite{ref1} on several key aspects. We expanded our evaluation to 15 different LLMs instead of only 3 (Llama2-70b, GPT-4, GPT-3.5-turbo) to gain a broader understanding of LLM's capabilities in materials science. Additionally, we chose not to include the chain-of-thought prompting method as preliminary results in~\cite{ref1} indicated that it did not significantly influence the performance of LLMs in answering materials science related questions. Another important difference came from the temperature parameter that regulates the stochasticity of LLMs response. Zaki~\textit{et al.} used a temperature of 1 during LLM's evaluations which allows for more randomness in the model's responses. However, we opted to use a temperature of 0 to ensure maximum determinism and consistency in the answers. A temperature of 0 ensures that a model chooses a most probable answer and provides a fairer assessment of models' knowledge integration and usage abilities. Indeed, the shape of the posterior distribution of tokens for a given input sequence being unknown for every LLMs, this would impose to propose two strategies for a fair evaluation: (i) Fix the temperature as we did, or (ii) find the best temperature for each LLM. The second strategy being costly and time-prohibitive, we opted for the first one such that the most probable output from each LLM are compared. 
    To also ensure the reliability of our results, we submitted each question to the models three times to assess the repeatability of their answers. Indeed, even though a temperature of 0 was fixed to maximize determinism in answers, uncontrollable features leading to stochasticity still remains such as floating-point precision~\cite{courbariaux2015low}, expert selection in mixture of experts (MoE) models like GPT-4 and Mixtral-8x7B~\cite{jiang2024mixtral}, multi-threaded operations, random number generator state differences between runs, etc~\cite{brugger2014parallel}. 
    
    Finally, we maintained consistency with the original study by using a same assistant prompt preceding every questions and instructing LLM's desired behaviour: "Solve the following question. Write the correct answer inside a list at the end." This approach allowed for direct comparison of our results to Zaki~\textit{et al.}~\cite{ref1}.

\begin{figure}[h]
\centering
  \includegraphics[width=\columnwidth]{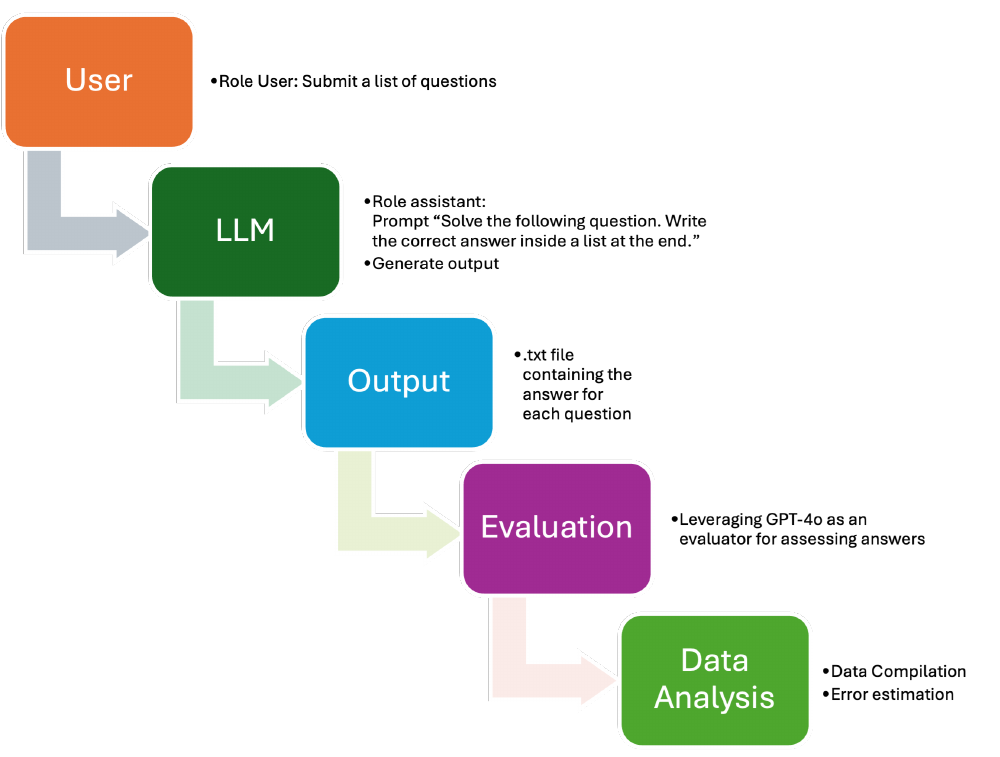}
  \caption{Pipeline for generating and evaluating responses from LLMs to the MaScQA benchmark.}
  \label{fgr:architecture}
\end{figure}

    We used the OpenAI, Anthropic and Ollama APIs to access the models~\cite{openai2023api, anthropic2023api, ollama2023api}. The models used in this study are: GPT-4-turbo, GPT-4o, GPT-4o-mini, GPT-4, GPT-3.5-turbo, Claude-3-Opus, Claude-3-Haiku, Claude-3.5-Sonnet, Llama2-7b, Llama2-70b, Llama3-8b, Llama3-70b, Mistral-7b, Phi3-3.8b and Phi3-14b. The tokenization process for all LLMs was handled automatically by the respective Python libraries, Ollama and OpenAI, which provide built-in tokenization as part of their APIs. No custom tokenization was applied in this study. Readers interested in the specifics of tokenization can refer to the official documentation of these libraries. The results were saved in *.txt files and are available on GitHub: \url{https://github.com/Lambard-ML-Team/LLM_comparison_4MS}. \\
    The LLMs were tested on two different machines: a MacBook Pro M1 (2020, 8GB RAM) and a GPU server ($\rm 8 \times $A100 40 GB PCIe NVIDIA GPUs). To assess the impact of hardware on performance only GPT-3.5-turbo, GPT-4, Llama2-7b, Llama3-8b have been tested on both machines. For models such as GPT-3.5-turbo and GPT-4 which only rely on OpenAI's servers, results remained consistent across both machines. However, for models like Llama2-7b and Llama3-8b, which run locally and are directly impacted by the host machine's specifications, performance variations were observed. Llama2-7b performed similarly on both machines, while Llama3-8b exhibited a 16\% performance improvement on the GPU server. To ensure optimal testing conditions, we divided the models based on their computational requirements and on machines' availability. The distribution of models is as follows: 
    
    \begin{itemize}
    \item \textbf{MacBook Pro M1:}
        GPT-4-turbo, GPT-4o, GPT-4, GPT-3.5-turbo, Claude-3-Opus, Claude-3-Haiku, Claude-3.5-Sonnet, Llama2-7b, Llama3-8b.
    \item \textbf{GPU server:}
        GPT-4, GPT-4o-mini, GPT-3.5-turbo, Llama2-7b, Llama2-70b, Llama3-8b, Llama3-70b, Mistral-7b, Phi3-3.8b, Phi3-14b.  
    \end{itemize}
    
    This distribution ensures that local models benefit from the GPU server's superior computational resources, providing a more accurate assessment of LLMs' capabilities under optimal conditions.
    In the study conducted in~\cite{ref1}, the evaluation of the LLMs' responses was manually performed. However, our study involves a significantly larger amount of LLMs' responses to evaluate, 19 LLMs (15 unique models and 4 models assessed on both machines) across three iterations for each of 644 questions, resulting in a total of $\sim$37,000 answers. Given the large scale of this dataset, manual evaluation would be impractical. Therefore, we applied a LLM-as-a-judge strategy~\cite{zheng2023judging} assisted by GPT-4o to handle this extensive volume efficiently and ensure accuracy. Figure~\ref{fgr:architecture} summarises the entire pipeline for generating answers and evaluating them. 
    
\subsection{Autonomous answers analysis}
\begin{figure*}[h]
\centering
  \includegraphics[width=\textwidth]{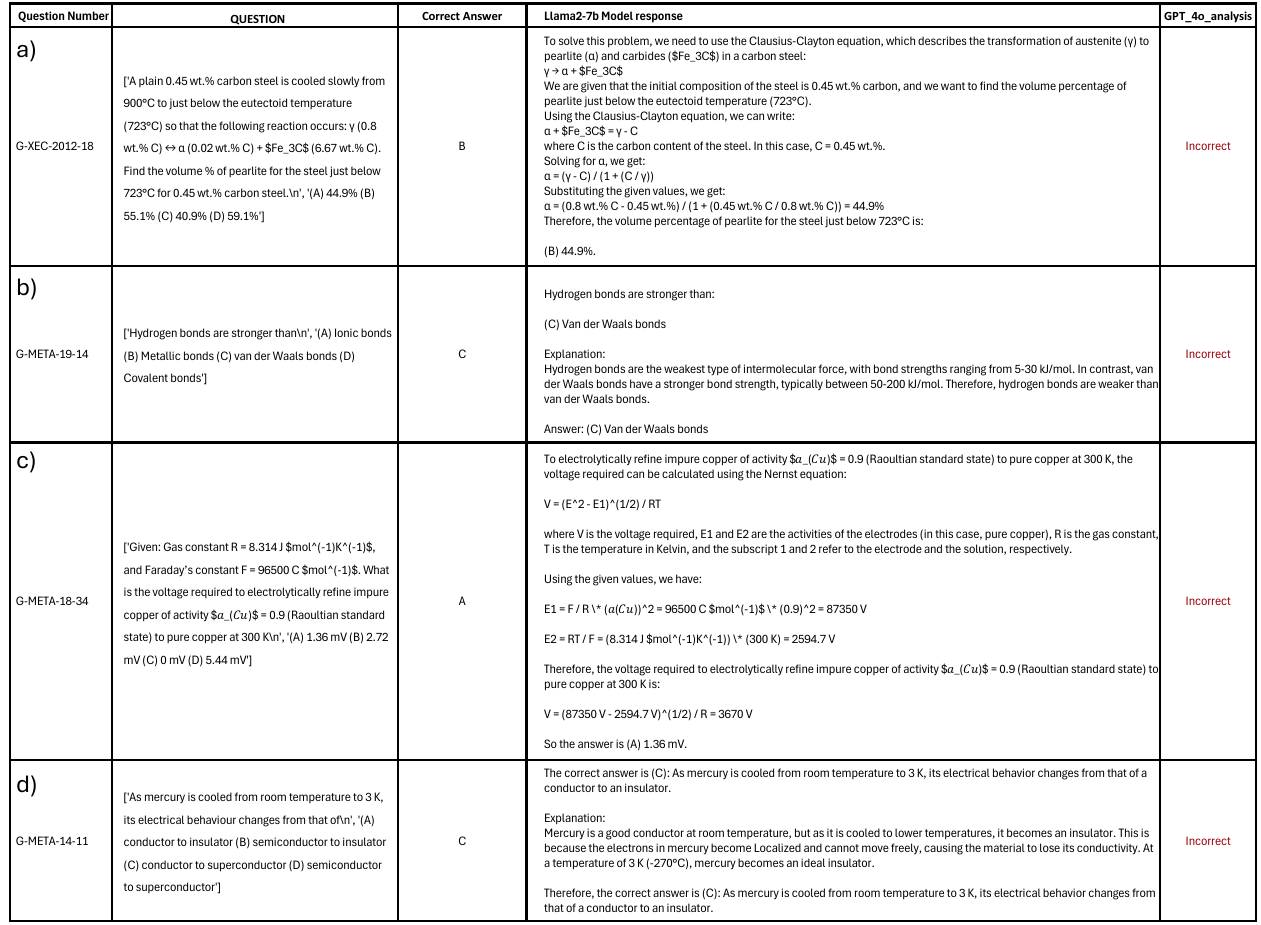}
  \caption{Example of ambiguous answers from Llama2-7b model analysed by GPT-4o. a) Wrong reasoning and calculation, selected the correct letter but associate the wrong value to it, b) Select the good answer but the reasoning says the opposite, c) Reasoning and calculation incorrect but selected the correct answer, d) select the correct answer but the reasoning and the text associated with the letter C is incorrect. }
  \label{fgr:example}
\end{figure*}

    To estimate the accuracy of GPT-4o to autonomously analyse LLMs' responses, we manually checked the results for four different LLMs. The manual analysis wasn't straightforward as certain models, mainly Llama2 and Llama3, provided ambiguous answers as shown in Figure~\ref{fgr:example}. Our approach to determining the correctness of these answers involved adopting the perspective of an examiner and evaluating whether the LLM's response matched the correct answer, focusing solely on the correctness of the selected option rather than the accompanying reasoning or explanatory text.

    As shown in Figure~\ref{fgr:example}, there are several types of ambiguous answers from the Llama2-7b model. Figure~\ref{fgr:example}(a) illustrates a case where the reasoning and calculation are incorrect, but the correct letter is selected with an incorrect value association. Figure~\ref{fgr:example}(b) shows the model selecting the correct answer while providing contradictory reasoning. Figure~\ref{fgr:example}(c) demonstrates a situation where the reasoning and calculation are incorrect, yet the correct answer is chosen. Finally, Figure \ref{fgr:example}(d) depicts the correct answer being selected despite incorrect reasoning and associated text.

    In the case of MATCH, MCQ, and MCQN questions, responses are assessed solely based on the selected letter (A, B, C, or D) rather than the accompanying reasoning, calculations, or explanatory text. Consequently, for such questions, the answers depicted in Figure \ref{fgr:example} should be considered correct if they align with the expected answer's letter, regardless of any associated reasoning or textual explanations.

    Finally, to validate GPT-4o's role as an evaluator, we performed a manual comparison of its judgments against human-assigned scores, as shown in Tables~\ref{tbl:misclassification} and~\ref{tbl:misclassification_v2}. This analysis demonstrates GPT-4o's accuracy as a judge while also identifying areas where discrepancies arise, particularly for questions requiring nuanced reasoning.
    
\subsubsection{First approach~~}

    Initially, we selected GPT-4o for this task, using a straightforward prompt: "Based on the question and the correct answer, You must tell if the other answer is correct or not by answering only with Correct or Incorrect" as shown in Figure~\ref{fgr:evaluation} a) and then submitted the question in the format: "The question is " + <QUESTION> + ", the correct answer is "+ <CORRECT ANSWER> + "and the other answer is : " + <MODEL ANSWER>. Consequently, the accuracy of GPT-4o in properly evaluating LLMs' answers was estimated to be an overall $\rm \sim 95.5$\% which is a strong performance, as shown in Table~\ref{tbl:misclassification}. We define here by "misclassification" the correct answers labelled as incorrect, and incorrect answers labelled as correct. However, we observed significant variation depending on the specific model being evaluated. Models with generally lower performance such as Llama2 and Llama3 were more susceptible to errors in the evaluation process. Notably, these models frequently had correct answers misclassified as incorrect more often than incorrect answers were misclassified as correct. For instance, Llama2-7b initially demonstrated 85/644 correct answers, however, we observed 48 correct answers as misclassified by GPT-4o. Despite maintaining an accuracy of $\rm \sim 92.5$\%, this misclassification resulted in Llama2-7b having an increase to 133 correct answers in total, reflecting a difference of $\rm \sim 56.5$\%.

\begin{table}[h]
\small
  \caption{\ Number of misclassification (over 644 questions) and estimated accuracy of the evaluating model GPT-4o with the first approach.}
  \label{tbl:misclassification}
  \begin{tabular*}{0.48\textwidth}{@{\extracolsep{\fill}}lcl}
    \toprule
    \textbf{Models}             & \textbf{\shortstack{Errors \\ GPT-4o}} & \textbf{\shortstack{Accuracy \\ GPT-4o}}\\
    \midrule
    Claude-3.5-Sonnet           & 10 & 98.4\% \\
    GPT-4-turbo                 & 17 & 97.4\%  \\
    Llama3-8b (MAC)             & 40 & 93.8\%  \\
    Llama2-7b (GPU server)           & 48 & 92.5\% \\
    \midrule
    \textbf{Overall Accuracy}   & - & \textbf{95.5\%}  \\
    \bottomrule
  \end{tabular*}
\end{table}
    
\subsubsection{Second approach~~}

\begin{figure*}[!h]
\centering
  \includegraphics[width=\textwidth]{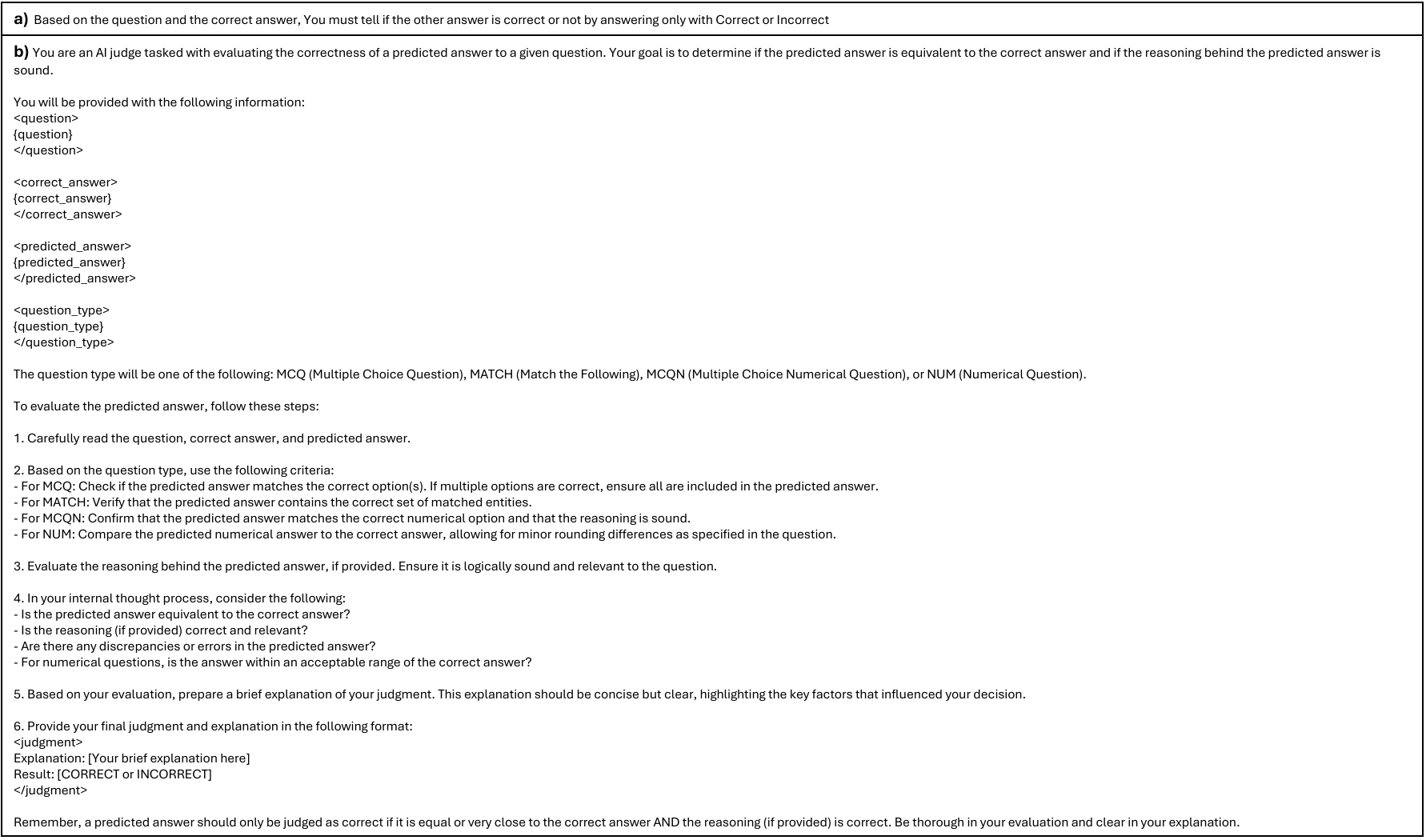}
  \caption{Comparison of the prompt used for the evaluation on GPT-4o: a) corresponds to the first approach with a straightforward prompt, while b) corresponds to the second approach with a step-by-step protocol and detailed explanation required.}
  \label{fgr:evaluation}
\end{figure*}

    In an attempt to resolve the issue of misclassification with the first approach, we decided to update the prompt for the evaluation to a more sophisticated one. In this new approach, the questions were formatted differently and the prompt described more precisely the task that GPT-4o had to perform. Specifically, the prompt instructed the model to evaluate not only the accuracy of the predicted answer but also the validity of the reasoning behind it, if provided. The model was required to ensure that the predicted answer matched the correct option, contained the correct set of matched entities, or was numerically accurate within an acceptable range. Furthermore, the model was tasked with providing a clear and concise explanation of its judgment, focusing on the key factors that influenced its decision. This refined prompt, shown in Figure~\ref{fgr:evaluation} b), enhanced the model's ability to interpret and evaluate answers more effectively, ultimately improving the accuracy and reliability of the evaluation process. 

\begin{table}[h]
\small
  \caption{\ Number of misclassification (out of 644 questions) and the corresponding estimated accuracy of the evaluating model GPT-4o when applying the second approach. The table also includes a comparative analysis with GPT-4o-mini.}
  \label{tbl:misclassification_v2}
  \begin{tabular*}{0.48\textwidth}{@{\extracolsep{\fill}}llll}
    \toprule
    \textbf{Models}             & \textbf{\shortstack{Errors \\ GPT-4o}} & \textbf{\shortstack{Accuracy \\ GPT-4o}} & \textbf{\shortstack{Errors \\ GPT-4o-mini}}\\
    \midrule
    Llama2-7b (GPU server)           & 15 & 97.7\% & 28 \\
    Llama3-8b (GPU server)           & 11 & 98.3\% & - \\
    Mistral-7b (GPU server)          & 16 & 97.5\% & - \\
    GPT-4 (GPU server)               & 11 & 98.3\% & 41 \\
    \midrule
    \textbf{Overall Accuracy}    & - & \textbf{97.9\%} & \textbf{94.6\%}  \\
    \bottomrule
  \end{tabular*}
\end{table}

As shown in Table~\ref{tbl:misclassification_v2}, the accuracy of the evaluation reached $\rm \sim97.9$\%, demonstrating a greater stability across different models. Notably, Llama2-7b's misclassifications decreased from 48 in the initial approach to 15, and Llama3-8b's misclassifications dropped from 40 to 11. This significant decrease in misclassifications highlights the effectiveness of the revised evaluation prompt. However, if the revised prompt is applied to the GPT-4o-mini model as a judge, the results were less conclusive when compared to GPT-4o, with 28 misclassifications observed for Llama2-7b and 41 for GPT-4. Historically, the model GPT-4o-mini was made available to the public by OpenAI during the evaluation process of the LLMs' answers, and its more attractive price tag conducted us to try it out on the benchmark. 

A key issue with GPT-4o-mini was its failure to recognize some correct answers when the evaluated LLM neglected to include the corresponding letter in its responses. This suggests that while the new prompt greatly enhances evaluation accuracy for higher-performing models, it may still be prone to errors with LLMs with lower reasoning capabilities or when critical elements, such as the letter designation in answers, are omitted. Future work could explore refining the prompt further to handle such cases more effectively or developing additional layers of validation to ensure even greater accuracy and consistency across all model types.

\subsection{Random baseline calculations}
\label{subsec:rbc}
To assess the extent to which LLMs outperform chance-level guessing, we compute a random baseline for each of the question types in the MaScQA benchmark. Knowing that each of the MATCH, MCQ, and MCQN questions has four options, with one correct answer, we derive the mean, \(\mu\), and standard deviation, \(\sigma\), of the expected number of correct answers from the properties of the binomial distribution, which models the number of successes (correct answers) in a fixed number of independent trials (questions), each with a fixed probability of success \( p = 0.25 \). Specifically:
\[ \mu = n \cdot p, \quad \sigma = \sqrt{n \cdot p \cdot (1-p)}, \]
where \( n \) is the number of questions for a given category, and \( p \) is the probability of guessing correctly.

Therefore, and as reported in Table~\ref{tbl:result}, we have:

\begin{itemize}
    \item For MATCH questions (70 total):
    \[ \mu = 70 \cdot 0.25 \approx 17.5, \quad \sigma = \left( 70 \cdot 0.25 \cdot 0.75 \right)^{0.5} \approx 3.6. \]
    \item For MCQ questions (283 total):
    \[ \mu = 283 \cdot 0.25 \approx 70.7, \quad \sigma = \left( 283 \cdot 0.25 \cdot 0.75 \right)^{0.5} \approx 7.3. \]
    \item For MCQN questions (67 total):
    \[ \mu = 67 \cdot 0.25 \approx 16.7, \quad \sigma = \left( 67 \cdot 0.25 \cdot 0.75 \right)^{0.5} \approx 3.5. \]
\end{itemize}

For NUM questions (224 total), a precise numerical reasoning is required, and the answers aren't multiple-choice. Thus, the probability of guessing correctly by chance is effectively close to zero. This stems from the nature of the problem: without predefined options, the likelihood of randomly selecting the correct answer in a continuous or large discrete range (e.g., all real numbers or integers) is negligible. Consequently, we fix the mean baseline accuracy for NUM questions at 0\% with equivalently 0\% in standard deviation, acknowledging the unlikelihood of finding the correct answer randomly on a continuous range of real numbers.

Finally, the combined \(\mu = 105.0\) and \(\sigma \approx 8.9\) for the entire set of MATCH, MCQ, MCQN, and NUM questions are derived from the sum of the means and variances (\(\sigma^2\)) of each question category, respectively. 

Thus, we can compare the performance of each LLM against this random baseline to highlight their ability for knowledge retrieval, logical reasoning, and numerical computation effectively.

\section{Results}

After establishing the accuracy of the methodology for the autonomous evaluation pipeline, the entire list of LLMs from Table~\ref{tbl:Information} were evaluated on the MaScQA benchmark with the results presented in Tables \ref{tbl:result} and \ref{tbl:result_percent}. Table \ref{tbl:result} summarizes the average correctness of each LLM across three iterations on the 644 benchmark questions. Additionally, to assess the impact of hardware on model performance, four LLMs (GPT-4, GPT-3.5-turbo, Llama2-7b and Llama3-8b) were tested on a MAC and a GPU server. This comparative evaluation offers valuable insights into how computational resources can influence the performance and accuracy of LLMs' responses. For GPT-4 and GPT-3.5-turbo, no performance differences were observed, as these models rely on the server infrastructure provided by OpenAI, thereby rendering the local hardware inconsequential. However, a notable performance increase of $\rm \sim 16$\% was observed for Llama3-8b when run on the GPU server in comparison to the MAC M1. Conversely, Llama2-7b showed no significant performance difference between the two machines, likely due to the MAC M1’s sufficient capability to handle the model effectively.

This disparity in performance, particularly with Llama3-8b, can be attributed to the computational demands exceeding the MAC M1’s capacity, whereas the GPU server, with superior hardware capabilities, could manage the workload without compromise. Additionally, when running Llama2-7b and Llama3-8b on the MAC M1, the system resources were fully utilized, leaving the machine unable to perform other tasks until completion. This was not the case on the GPU server, where system performance remained stable, underscoring the importance of hardware resources in managing complex models like Llama3-8b.

\begin{figure}[!h]
  \includegraphics[width=\columnwidth]{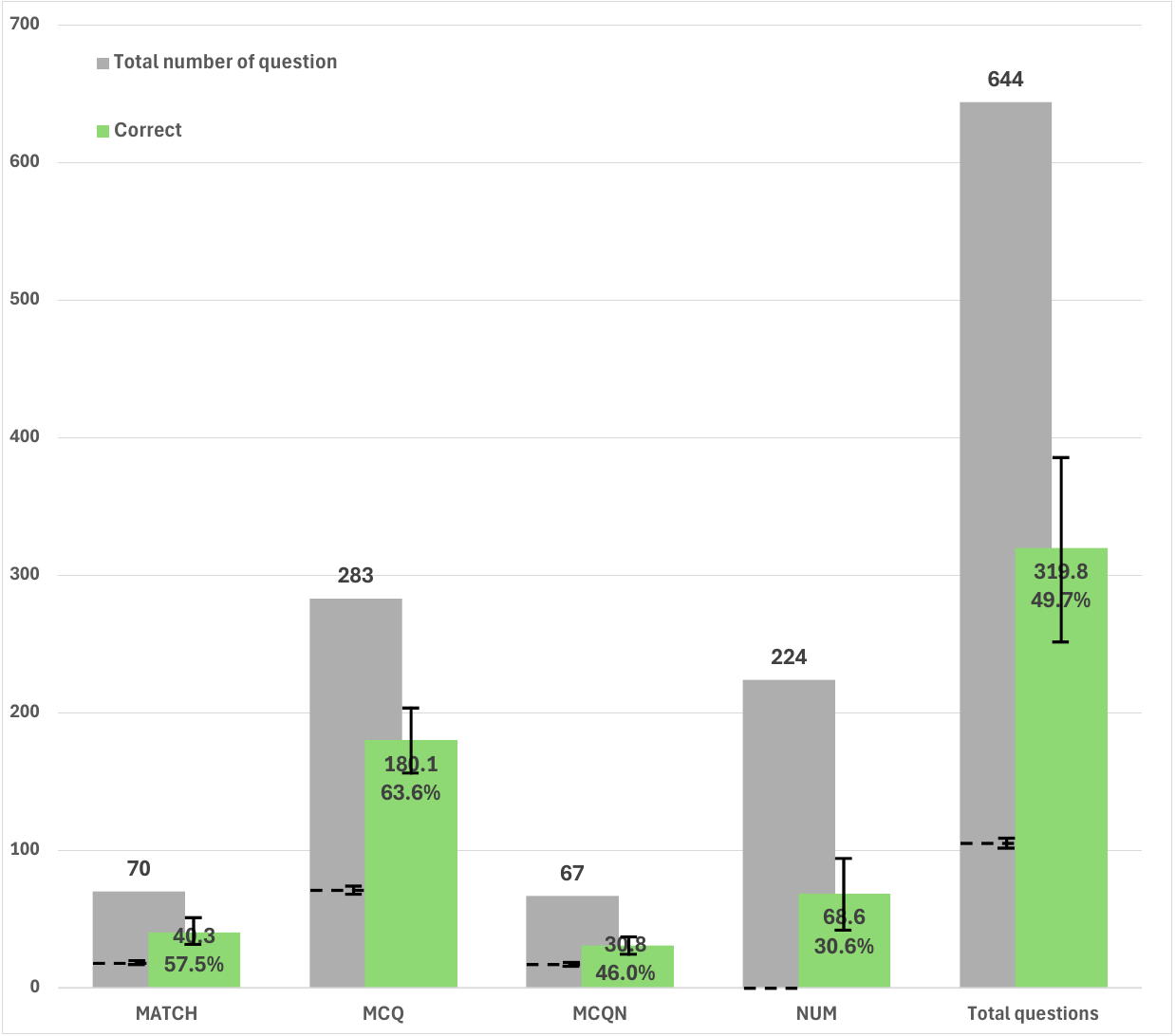}
  \caption{Comparison of the number of average correct answers, including the 15 unique LLMs tested, to the total number of questions per category, i.e., MATCH, MCQ, MCQN, and NUM, as well as for the whole set of questions. A random baseline per category is indicated as a dashed line.}
  \label{fgr:average}
\end{figure}

Figure~\ref{fgr:average} illustrates that, in general, LLMs tend to demonstrate higher accuracy when responding to questions that provide a set of possible answers (MATCH, MCQ and MCQN).  This phenomenon can be explained by the fact that, for the type of questions with multiple choices available, the model is required to select from a predefined list of options. Similar to a student guessing the correct answer, the model may choose the correct option even if the underlying reasoning or calculations are flawed. This tendency is further demonstrated in Figure~\ref{fgr:example}, where models exhibited correct selections despite incorrect reasoning.

An important aspect of our analysis is the evaluation of the LLMs on NUM, which present a unique challenge as they do not provide potential answers. This type of question requires models to rely solely on their internal knowledge, reasoning, and computational abilities. The results for NUM, as depicted in Table~\ref{tbl:result_percent}, offer a clear depiction of the LLMs' capabilities in these areas. Notably, the performance of the models on NUM questions reveals distinct groups. 
The difficulties observed in MaScQA’s NUM and MCQN categories align with challenges reported in benchmarks such as MATH~\cite{hendrycks2021measuringmathematicalproblemsolving} and ChemBench4k~\cite{zhang2024chemllm}. These tasks often require multi-step computations, reasoning under constraints, and precision in numerical outputs—areas where current LLMs frequently fall short.

Models like Llama2-7b and Mistral-7b, which performed worse than random in MCQN, highlight a persistent issue of shallow numerical reasoning and tokenization inefficiencies. Addressing these limitations may require targeted fine-tuning with domain-specific datasets or improved model architectures better suited for handling numerical reasoning tasks.

As shown in Tables~\ref{tbl:result} and~\ref{tbl:result_percent}, and Figure~\ref{fgr:average}, most of tested LLMs outperform in average the random baseline in all question categories, except for Llama2-7b in the MATCH and MCQN categories, as well as for Mistral-7b in the MCQN category. For those two last LLMs, their results in the MCQN category seem to be hindered by their poor capability on numerical computations, as their performance on the MCQ category alone outperforms the random baseline. However, concerning the behavior of Llama2-7b in the MATCH category, it could imply that Llama2-7b rather follow systematic flawed reasoning patterns learned from its training data that aren't fitted to materials science and engineering. Additionally, the lack of domain-specific knowledge is hypothesized to also be a culprit. This emphasizes the need for domain-targeted fine-tuning or retraining to align LLMs with materials science tasks. Importantly, such behavior underscores the value of rigorous benchmarking across diverse question types to identify and address weaknesses in model reasoning capabilities.
Also, issues observed in MATCH and MCQ categories are not unique to MaScQA. Similar limitations have been identified in benchmarks like SciQ~\cite{SciQ} and MoleculeQA~\cite{lu2024moleculeqa}. For MATCH tasks, LLMs struggle to establish logical relationships between entities, often defaulting to heuristic-based reasoning. MCQ tasks, while simpler, can be impacted by pattern exploitation where models rely on superficial cues rather than true conceptual understanding.

These trends underscore the importance of prompt optimization and domain-specific fine-tuning to improve structured reasoning and conceptual alignment in materials science tasks. Future work could explore methods to guide models more effectively through MATCH-type reasoning frameworks and numerical computations.

Claude-3.5-Sonnet emerges as the top performer, closely followed by GPT-4o, both achieving an accuracy exceeding $\rm \sim 70$\%. This level of accuracy is considered acceptable given the complexity of the task. Claude-3-Opus and GPT-4-turbo closely follow with $\rm \sim 64-63$\%, both models demonstrating a large effectiveness at handling numerical computations by comparison to the average pool of LLMs topping at $\rm \sim 30.6$\% (see Figure~\ref{fgr:average}). Notably, the best studied open-source model, Llama3-70b, achieves results that are closely aligned with those of GPT-4 and Claude-3-Haiku with $\rm \sim 32.6$\%, underscoring its competitiveness with closed-source models.

Furthermore, the performance comparison between Phi3-3.8b, Phi3-14b, and GPT-3.5-turbo reveals minimal differences, suggesting that the parameter count may not be the sole determinant of a LLM's effectiveness. Interestingly, Phi3-3.8b outperforms several models with double its parameter count, including Llama3-8b, Mistral-7b, and Llama2-7b. The relatively poor performance of these larger models highlights the complexity of balancing model size with other factors such as architecture and training data quality, which can significantly impact overall performance.

\begin{table*}[h]
    \centering
    \caption{\ Number of correct answers achieved by 19 Large Language Models (LLMs) (representing 15 unique models) on the MaScQA~\cite{ref1} benchmark. Each model was evaluated through three submissions per question to ensure robustness and consistency of results. Some LLMs were tested on two different machines to assess potential variations in performance. The numbers in parentheses indicate the number of questions within each category. For comparison, we also incorporate a random baseline as computed in section~\ref{subsec:rbc}.}
      \label{tbl:result}
    \begin{tabular*}{\textwidth}{@{\extracolsep{\fill}}lllllll}
        \toprule
        Machine used                & LLM                   & MATCH (70) & MCQ (283) & MCQN (67) & NUM (224) & \shortstack{Total correct \\answer (644)} \\
        \midrule
        \multirow{9}{*}{Mac Pro M1} & GPT-4-turbo           & 65.0±1.0 & 236.8±2.8 & 48.8±2.7 & 141.2±3.5 & 491.8±4.5 \\
                                    & GPT-4o                & 67.9±0.9 & \textbf{260.1±2.2} & 50.7±2.0 & 161.0±5.9 & 539.7±8.2 \\
                                    & GPT-4                 & 60.4±1.4 & 214.8±2.4 & 34.4±0.2 & 80.4±6.9 & 390.1±3.5 \\
                                    & GPT-3.5-turbo         & 25.1±2.7 & 157.8±2.2 & 29.1±1.3 & 47.8±3.7 & 259.8±8.4 \\
                                    & Claude-3-Opus         & 68.7±0.6 & 240.3±0.6 & 49.2±0.2 & 143.6±3.8 & 501.8±3.7 \\
                                    & Claude-3-Haiku        & 40.3±0.6 & 205.1±0.2 & 33.0±0.3 & 77.0±0.3 & 355.4±0.5 \\
                                    & Claude-3.5-Sonnet     & \textbf{69.0±0.0} & 248.8±0.7 & \textbf{55.1±2.0} & \textbf{167.1±0.2} & \textbf{540.0±1.3} \\
                                    & Llama2-7b             & 9.3±2.4 & 99.2±1.6 & 14.7±4.8 & 5.8±1.7 & 129.0±4.9 \\
                                    & Llama3-8b             & 22.5±0.7 & 132.9±1.1 & 15.1±0.8 & 18.2±1.1 & 188.8±1.2 \\
        \midrule
        \multirow{ 10}{*}{GPU server}    & GPT-4                 & 61.4±0.5 & 212.4±2.7 & 33.9±1.7 & 85.7±2.3 & 393.4±3.6 \\
                                    & GPT-4o-mini           & 59.2±0.4 & 226.9±1.1 & 47.1±0.9 & 120.8±3.3 & 454.0±4.6 \\
                                    & GPT-3.5-turbo         & 24.0±3.6 & 158.3±1.5 & 30.0±3.2 & 49.9±0.5 & 262.2±0.9 \\
                                    & Llama2-7b             & 9.1±3.7 & 98.9±10.1 & 12.3±2.8 & 5.0±2.9 & 125.3±10.4 \\
                                    & Llama2-70b            & 18.9±3.6 & 129.3±4.1 & 20.7±3.0 & 11.8±0.7 & 180.7±8.7 \\
                                    & Llama3-8b             & 21.5±4.2 & 153.8±1.1 & 22.8±4.1 & 21.1±0.9 & 219.1±5.1 \\
                                    & Llama3-70b            & 51.8±0.9 & 199.2±2.5 & 36.5±2.0 & 73.0±3.6 & 360.6±1.9 \\
                                    & Mistral-7b            & 19.4±2.9 & 129.2±5.2 & 10.0±2.9 & 14.4±5.1 & 173.1±6.8 \\
                                    & Phi3-3.8b             & 32.9±1.4 & 146.8±3.9 & 18.8±1.0 & 36.8±6.1 & 235.2±9.6 \\
                                    & Phi3-14b              & 38.5±3.5 & 170.5±5.0 & 23.9±3.6 & 43.0±5.4 & 275.8±7.4 \\
        \bottomrule
        Random baseline    & --- & 17.5±3.6 & 70.7±7.3 & 16.7±3.5 & 0.0±0.0 & 105.0±8.9 \\
        \bottomrule
    \end{tabular*}
\end{table*}

\begin{table*}[h]
  \caption{Performance (accuracy (\%)) for 15 different LLMs evaluated on the MaScQA~\cite{ref1} benchmark. Each LLM was assessed through three submissions for each question to ensure robustness and consistency of results. For comparison, we also incorporate a random baseline as computed in section~\ref{subsec:rbc}.}
  \label{tbl:result_percent}
  \begin{tabular*}{\textwidth}{@{\extracolsep{\fill}}llllll}
    \toprule
    \textbf{Models}     & \textbf{MATCH (\%)} & \textbf{MCQ (\%)} & \textbf{MCQN (\%)} & \textbf{NUM (\%)} & \textbf{\shortstack{Overall \\ Accuracy}(\%)}\\
    \midrule
    Claude-3-Haiku      & 57.6±0.8 & 72.5±0.1 & 49.3±0.4 & 34.4±0.1 & 55.2±0.1 \\
    Claude-3-Opus       & 98.1±0.8 & 84.9±0.2 & 73.4±0.3 & 64.1±1.7 & 77.9±0.6 \\ 
    Claude-3.5-Sonnet   & \textbf{98.6±0.0} & 87.9±0.2 & \textbf{82.2±3.0} & \textbf{74.6±0.1} & \textbf{83.9±0.2 }\\
    GPT-3.5-turbo       & 35.1±4.2 & 55.9±0.6 & 44.1±3.3 & 21.8±1.2 & 40.5±0.9 \\
    GPT-4               & 87.0±1.6 & 75.5±0.9 & 51.0±1.7 & 37.1±2.4 & 60.8±0.6 \\
    GPT-4-turbo         & 92.9±1.4 & 83.7±1.0 & 72.8±4.1 & 63.0±1.6 & 76.4±0.7 \\
    GPT-4o              & 97.0±1.2 & \textbf{91.9±0.8} & 75.6±3.0 & 71.9±2.6 & 83.8±1.3 \\
    GPT-4o-mini         & 84.6±0.6 & 80.2±0.4 & 70.3±1.3 & 53.9±1.5 & 70.5±0.7 \\
    Llama2-7b           & 13.2±4.0 & 35.0±2.3 & 20.1±5.6 & 2.4±1.0 & 19.7±1.2 \\
    Llama2-70b          & 27.0±5.2 & 45.7±1.4 & 30.8±4.4 & 5.3±0.3 & 28.1±1.4 \\
    Llama3-8b           & 31.4±3.9 & 50.6±4.1 & 28.3±7.4 & 8.8±0.8 & 31.7±2.6 \\
    Llama3-70b          & 74.0±1.2 & 70.4±0.9 & 54.5±2.9 & 32.6±1.6 & 56.0±0.3 \\
    Mistral-7b          & 27.8±4.1 & 45.7±1.8 & 14.9±4.3 & 6.4±2.3 & 26.9±1.0 \\
    Phi3-3.8b           & 47.0±2.0 & 51.9±1.4 & 28.1±1.5 & 16.4±2.7 & 36.5±1.5 \\
    Phi3-14b            & 55.0±5.0 & 60.2±1.8 & 35.7±5.3 & 19.2±2.4 & 42.8±1.1 \\
    \bottomrule
    Random baseline    & 25.0±5.2 & 25.0±2.6 & 25.0±5.3 & 0.0±0.0 & 16.3±1.4 \\
    \bottomrule
  \end{tabular*}
\end{table*}

\begin{figure*}[h]
  \includegraphics[width=\textwidth]{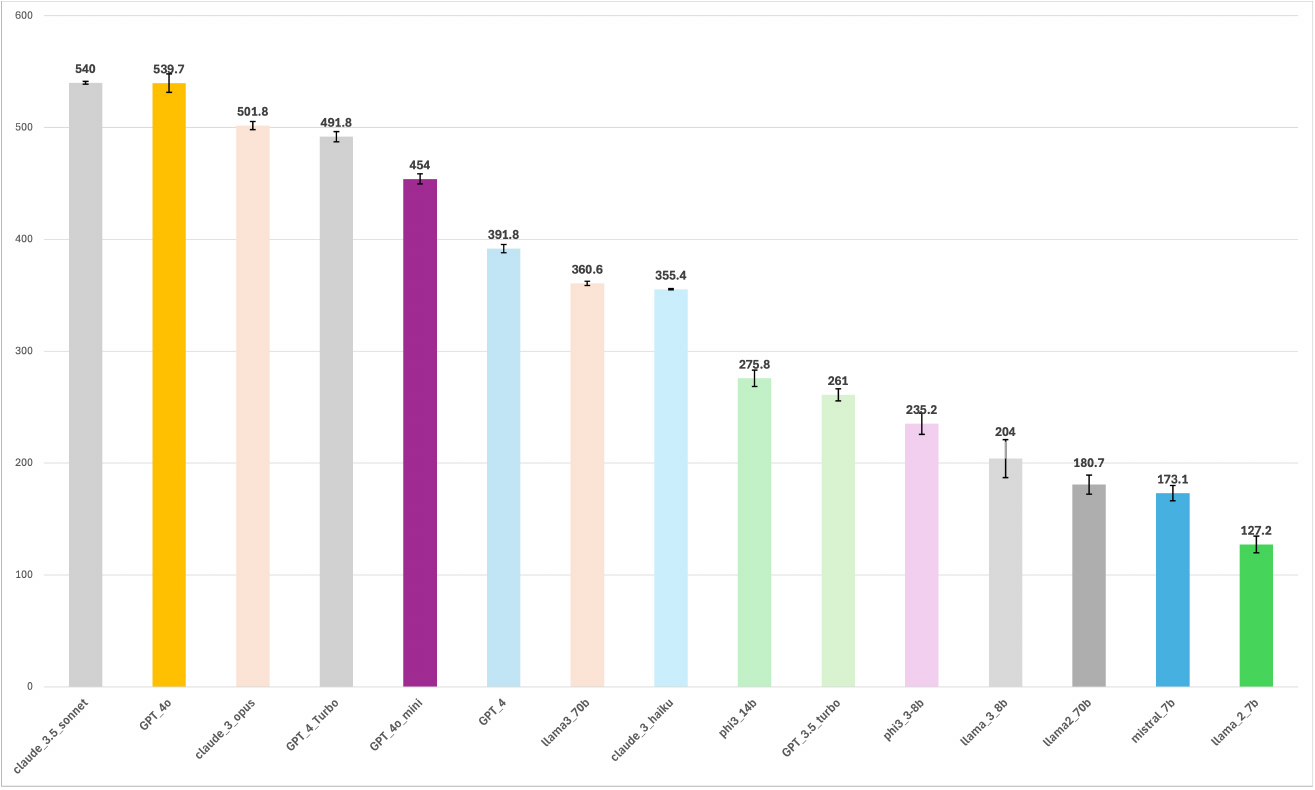}
  \caption{Average overall performance for the studied 15 unique LLMs with their standard deviation obtained from three runs over the whole set of 644 MATCH, MCQ, MCQN, and NUM questions.}
  \label{fgr:global_eval}
\end{figure*}

The models utilized in the study by Zaki~\textit{et al.}~\cite{ref1} show comparable performance to those in our current study. Notably, Llama2-70b exhibited slightly improved performance in our evaluation, with an accuracy of $\rm 28.1 \pm 1.4$\% compared to the $\rm 24.0$\% reported by Zaki~\textit{et al.} This difference could be attributed to the application of the chain-of-thought (CoT) technique on Llama2-70b in their study, as well as the systematic variation in computational resources and machines used.

In contrast, GPT-4 and GPT-3.5-turbo demonstrated consistent performance across both studies. Specifically, GPT-4 achieved an accuracy of $\rm 60.8 \pm 0.6$\% in our work, closely aligning with the $\rm 61.38$\% reported by Zaki~\textit{et al.} Similarly, GPT-3.5-turbo performed at $\rm 40.5 \pm 0.9$\%, which is consistent with the $\rm 38.31$\% observed in their study. These results suggest that the performance of these models is robust across different experimental setups and conditions. The slight variations in accuracy can likely be attributed to the difference in temperature settings used during evaluation.

The evaluation of the LLMs, on Table~\ref{tbl:result_percent} and Figure~\ref{fgr:global_eval}, demonstrates that Claude-3.5-Sonnet and GPT-4o are among the top performers, achieving overall accuracies of approximately 84\% (see Figure 1 in the supplementary materials for details concerning the LLMs' average accuracy on each category MATCH, MCQ, MCQN, and NUM). Claude-3.5-Sonnet emerges as the highest performer, with an overall accuracy of 83.9\% with a high stability. Its exceptional performance across MATCH and NUM categories underscores its proficiency in pattern recognition and numerical reasoning, suggesting that it excels in tasks requiring both structured matching and complex calculations.
GPT-4o closely follows with an overall accuracy of 83.8\%. It demonstrates particular strength in the MCQ category, attaining the highest accuracy of 91.9\%. This indicates that GPT-4o is highly effective at handling multiple-choice questions where options are provided. Additionally, GPT-4o's performance in NUM at 71.9\% suggests a solid capability in numerical reasoning, although it slightly lags behind Claude-3.5-Sonnet in this area. 

Claude-3-Opus and GPT-4-turbo also exhibit commendable performance, with overall accuracies of 77.9\% and 76.4\%, respectively. These models show a balanced capability across different question types, reflecting their robustness and versatility in handling diverse tasks. Their relatively high performance across MATCH and MCQ categories indicates that they are reliable choices for a range of question types, though they do not quite reach the top levels achieved by Claude-3.5-Sonnet and GPT-4o.

GPT-4 and GPT-4o-mini achieved overall accuracies of 60.8\% and 70.5\%, respectively. While GPT-4 had lower performance in the NUM category, it was relatively strong in MATCH and MCQ categories.
Llama3-70b also falls into the mid-tier category with an overall accuracy of 56.0\%. Although it did not outperform the leading models, it showed decent performance in MATCH and MCQ categories. This model's performance highlights its capability in handling structured questions, although it still lags behind the top performers.
Llama2-7b, Llama2-70b, Llama3-8b, and Mistral-7b exhibited poor performance across all categories, with overall accuracies below 32\%. These models struggled particularly in the NUM category, where their accuracies were very low (ranging from 2.4\% to 8.8\%). This significant shortfall in numerical reasoning capabilities underscores major limitations in these models' ability to handle complex quantitative tasks, which might be due to their training data or architectural constraints. Also, several factors may explain the observed limitations of open-source models on numerical reasoning tasks:

\begin{itemize}
    \item Training data limitations: Open-source models are often trained on publicly available datasets, which may lack sufficient examples of numerical reasoning, particularly in scientific domains like materials science.
    \item Tokenization inefficiencies: Numbers are tokenized as sequences rather than atomic units, leading to errors in operations involving precision or formatting.
    \item Smaller model capacity: Models with fewer parameters have limited ability to perform complex, multi-step computations compared to larger closed-source counterparts.
    \item Reasoning biases: Open-source models prioritize fluency during pretraining, resulting in outputs that appear plausible but lack numerical accuracy.
\end{itemize}

Then, Phi3-3.8b and Phi3-14b performed better than the models explained before, with overall accuracies of 36.5\% and 42.8\%, respectively. Despite these improvements, their performance still fell short of the top-tier models, particularly in complex tasks such as MCQN and NUM. This suggests that while these models have some capabilities, they are not yet competitive with the leading models in handling more challenging question types.

Addressing these gaps requires a combination of strategies. For example, fine-tuning open-source models on curated datasets with extensive numerical tasks could significantly improve their reasoning capabilities. Additionally, advancements in tokenization strategies and enhanced pretraining methods could help smaller models better handle numerical precision, rounding, and formatting—critical elements for scientific applications like materials discovery.

Such targeted improvements are particularly relevant for tasks like calculating material properties or designing experiments, where numerical accuracy is essential. By bridging these gaps, open-source models can evolve into robust tools for domain-specific applications in materials science.

On the perspective of the categories of questions: 
\begin{itemize}
    \item MATCH: Claude-3.5-Sonnet achieved the highest accuracy (98.6\%), closely followed by Claude-3-Opus (98.1\%) and GPT-4o (97.0\%). This suggests these models are particularly adept at tasks requiring pattern recognition and matching. The high accuracy across these models suggests their robust capability in identifying and matching patterns effectively.
    \item MCQ: GPT-4o led in this category with a 91.9\% accuracy, indicating its strength in handling multiple-choice questions with provided options, reflecting its ability to navigate through choices efficiently.
    \item MCQN: Claude-3.5-Sonnet achieved an accuracy of 82.2\%, its capability to integrate numerical reasoning within the context of multiple-choice questions. The model’s strong performance in this category suggests it can effectively handle questions that require both choice selection and numerical computation.
    \item NUM: The NUM category, which requires open-ended numerical answers without provided options, was the most challenging. Claude-3.5-Sonnet performed best with a 74.6\% accuracy, its advanced numerical reasoning abilities suggests that it is particularly adept at generating accurate numerical responses when no options are provided.
\end{itemize}

The results in Figure~\ref{fgr:global_eval} highlight that while different models exhibit strengths in specific areas, Claude-3.5-Sonnet’s performance across both pattern recognition and numerical reasoning tasks positions it as a particularly versatile model. The challenges observed in the NUM category across all models underscore the need for continued advancements in handling open-ended numerical reasoning tasks.

\section{Discussion}

The results of this study underscore the current superiority of closed-source models, such as GPT and Claude families of models, over their open-source counterparts like Llama, Mistral, and Phi3. Closed-source models consistently demonstrated higher accuracy across various question categories, indicating their advanced architecture, extensive training, and optimization for a broad range of tasks, including the fields of materials science and engineering. However, the potential of open-source models should not be overlooked. Despite their lower performance in this benchmark, open-source models offer opportunities for optimization through methods like prompt engineering and fine-tuning. Fine-tuning, in particular, is a powerful tool that allows these models to be adapted to specific tasks or datasets, potentially enhancing their performance in specialized domains such as in materials science and chemistry. 

Overall, the inclusion of a random baseline for the MATCH, MCQ, MCQN, and NUM categories highlights the significant advantage provided by LLMs in answering materials science questions. For most of the tested LLMs, except Llama2-7b and Mistral-7b, they achieve accuracies demonstrating their ability to display a reasoning, i.e., a consistent arrangement of their fragments of memorized knowledge, and retrieve information far beyond chance-level guessing. Notably, the NUM category, which lacks predefined options, showcases the models' numerical reasoning capabilities---a critical skill for tasks such as calculating material properties or experimental parameters.

Phi3-3.8b stands out as a particularly promising candidate for such optimization. Despite having a relatively low number of parameters, it achieved an overall accuracy of 36.5\%, which is commendable given its smaller scale. This suggests that with targeted fine-tuning and prompt optimization, Phi3-3.8b could potentially improve its performance significantly without demanding an expensive hardware load. 

An interesting direction for future work could involve systematically fine-tuning Phi3-3.8b and other open-source models on domain-specific datasets, such as materials science or other technical fields. The MaScQA benchmark results directly inform the development of a RAG system tailored for materials science applications. Such a system will enable AI tools to assist researchers in tasks like synthesizing knowledge from massive literature corpora, proposing experimental designs, and predicting material properties with minimal human input.

For example, strong performance on NUM and MCQ questions demonstrates an LLM’s capability to accurately calculate material parameters or resolve conceptual queries—skills essential for automating computational tasks or pre-experimental analyses. Fine-tuning open-source models like Phi3-3.8b using curated materials science datasets will ensure these tools become domain-optimized, democratizing access to AI-powered solutions in materials research. Additionally, prompt engineering strategies could be explored to better leverage the model's existing capabilities, potentially boosting its performance in specific tasks. By carefully crafting prompts that guide the model's reasoning process, we can help it generate more accurate and contextually appropriate responses. This approach is particularly useful for numerical reasoning tasks, where precise wording can influence the model's output. These approach not only aim to bridge the performance gap between open- and closed-source models but also promote the democratization of AI by enhancing the utility of models that are freely accessible to the community.

While closed-source models currently lead in performance, the flexibility and accessibility of open-source models present a valuable opportunity for ongoing research and development. By focusing on fine-tuning and prompt optimization, it is possible to enhance the performance of open-source models, making them viable alternatives for specialized applications and contributing to the advancement of open AI technologies for diverse domains, materials science included.

While GPT-4o provides a creative and scalable approach for automating performance evaluation, it is not without limitations. Discrepancies between GPT-4o's assessments and human-assigned scores highlight challenges such as potential biases in LLM judgments, inconsistencies in reasoning, and difficulties with questions requiring deeper conceptual understanding. For this reason, we have complemented GPT-4o-based evaluations with traditional accuracy metrics, ensuring that the results remain quantitatively robust and reliable. Future work could explore hybrid evaluation frameworks that combine automated LLM-based scoring with rigorous manual validation.

The discrepancy observed in evaluation errors for lower-performing models suggests that outputs from these models are more challenging for automated evaluators like GPT-4o to assess accurately. Also, several factors could contribute to the higher susceptibility of lower-performing models to evaluation errors:

\begin{itemize}
    \item Ambiguity in outputs: Lower-performing models often produce ambiguous or incomplete answers, which are inherently harder to evaluate. Outputs may include partially correct information or lack the precision required, particularly for numerical and structured tasks.
    \item \textit{Hallucinations} and shallow reasoning: These models are more prone to hallucinations—confident but incorrect outputs—and rely on superficial reasoning, especially when confronted with multi-step or complex questions. Such outputs can mislead evaluators like GPT-4o.
    \item Tokenization and numerical precision issues: Numerical reasoning tasks (e.g., NUM) require strict handling of tokenization and precision. Lower-quality models frequently generate outputs with formatting errors or rounding inconsistencies, increasing evaluation discrepancies.
    \item Evaluator bias: Automated evaluators like GPT-4o may exhibit biases toward linguistic fluency and coherence. Outputs from lower-performing models, which tend to lack these qualities, can be disproportionately misclassified.
\end{itemize}

These observations offer a preliminary explanation for the observed phenomenon. A more detailed investigation involving model-level diagnostics or deeper access to closed-source architectures would be required to fully analyze this behavior. Future work could focus on developing error analysis frameworks and improving evaluator calibration to better handle outputs from lower-performing models.

This study represents a critical first step in identifying the best-performing LLMs as candidates for fine-tuning and integration into a materials science RAG system. To further advance the applicability of LLMs in materials science, several directions for future work are identified:

\begin{itemize} 
    \item Fine-tuning open-source models: While models like Phi3-3.8b show promise, fine-tuning on curated, domain-specific datasets rich in materials science literature and numerical reasoning tasks will be essential for improving their capabilities.
    \item Exploring temperature effects: Adjusting temperature settings dynamically could optimize model outputs for tasks requiring both creativity and precision, particularly in numerical and reasoning-heavy questions.
    \item Advanced error correction strategies: Implementing techniques such as CoT prompting, in-context learning (ICL), and post-hoc validation methods will address hallucinations, ambiguity, and shallow reasoning in lower-performing models.
    \item Improved tokenization for numerical tasks: Enhancing tokenization strategies to treat numerical inputs as atomic units rather than sequences will reduce errors in numerical reasoning and precision.
\end{itemize}

The end goal is to create an AI system capable of comprehensively reasoning over materials science knowledge, accelerating discoveries and reducing the time between hypothesis generation and experimental validation.

\section*{Conclusions}
This study used the MaScQA benchmark, developed by Zaki~\textit{et al.}~\cite{ref1}, to assess the performance of 15 different LLMs across a diverse set of tasks. The MaScQA dataset is notable for its inclusion of questions from various sub-fields from the materials science and engineering and its range of question types MATCH, MCQ, MCQN, and NUM, each of which evaluates different aspects of model capability, such as reasoning, pattern recognition, numerical computation, and decision-making. Among the models tested, two demonstrated exceptional performance: Claude-3.5-Sonnet and GPT-4o. Claude-3.5-Sonnet achieved an overall accuracy of 83.9±0.2\%, while GPT-4o closely followed with an accuracy of 83.8±1.3\%. These results highlight the advanced capabilities of these models in handling a wide array of tasks, particularly in domains requiring robust pattern recognition and complex numerical reasoning.

The variety of question types in the MaScQA benchmark allowed for a comprehensive evaluation of the LLMs, revealing not only the strengths of the top-performing models but also the specific areas where other models struggled. For instance, the NUM category, which involves open-ended numerical questions, proved to be particularly challenging for most models, underscoring the ongoing difficulties in developing LLMs with strong numerical computation abilities.

Overall, the findings from this study emphasize the potential of using benchmarks like MaScQA to push the boundaries of LLM capabilities for specific domains like materials science and engineering. The high performance of Claude-3.5-Sonnet and GPT-4o suggests that while state-of-the-art models continue to improve, there remains significant potential for further improvements, particularly for open-source models that can be fine-tuned and optimized for specific tasks. Future work in this area will focus on enhancing the capabilities of open-source models through targeted fine-tuning and prompt engineering, potentially narrowing the gap between open- and closed-source models and contributing to the broader development of accessible and high-performing AI systems for science. 

\section*{Author contributions}
Christophe Bajan: conceptualization, methodology, software, data analysis, writing --- original draft. Guillaume Lambard: conceptualization, methodology, software, validation, resources, supervision, funding acquisition, project administration, writing --- final draft. 

\section*{Conflicts of interest}
There are no conflicts to declare.

\section*{Data availability}

The code, dataset, raw responses from LLMs are accessible at the following URL: \url{https://github.com/Lambard-ML-Team/LLM_comparison_4MS} .




\balance


\bibliography{biblio}
\bibliographystyle{rsc}
\end{document}